m\magnification\magstep2
\overfullrule 0 pt
\baselineskip=18pt

\centerline{\bf On explicit solutions for completely integrable}
\centerline{\bf classical Calogero-Moser systems in external fields}

\vskip 0.5 in
\centerline{\bf D.V.Meshcheryakov, T.D.Meshcheryakova }
\vskip 0.2 in
\centerline{\it Physics Department, Moscow State University,}
\centerline{\it Leninskie gory, 119899, Moscow, Russia}

\vskip 0.5 in
\centerline{ Abstract }
Explicit solutions for one completely-integrable
system of Calogero-Moser type in external fields are found in case
of three and four interacting particles.  Relation between coupling
constant, initial values of coordinates and time of falling into the
singularity of potential is derived.

\vskip 1 in

\centerline{\bf 1. Introduction}

{\ \ \ }Completely integrable finite-dimensional many-particle
systems still attract considerable attention[1-6].
By now different methods of solving equations of motion have been
developed for many potentials giving the property of complete
integrability [1-3]. One of such methods for Calogero-Moser systems
in external fields was found in [5]. Here we apply this method in
some special cases of two-parameter Calogero-Moser type potential to
obtain explicit solutions of the equations of motion. We investigate
three and four particle systems. We analyze explicit solutions and
estimate the time of falling particles into the singularities of the
potentials for different initial conditions.

Complete integrability is guaranteed by the existence of $N$
involutory constants of motion. One of these constants is
Hamilton function which has the form

$$ H= \sum_{i=1}^{N} \left({p_i^2 \over 2} + W(x_i) \right)
+ \sum_{i>j}^N V(x_i -x_j) \eqno(1) $$

 System (1) is completely integrable when and only when the
external field potential $W(x)$ and pair interaction potential
$V(x)$ are the following:

$$ V(x)={ a \over
{x^2}},{\ \ }W(x)= \gamma_1 x^4+ \gamma_2 x^2 + \gamma_3 x \eqno(2)
$$ where $ a, \gamma_i, i=1,2,3 $ are arbitrary real constants.

In [5] a connection of N-particle system with Hamiltonian (1)
and potentials (2) with non-homogeneous Burgers--Hopf equation was
established.  This connection led to the method of solving
equations of motion for systems (1)-(2). If the constants defining
potentials satisfy the following constraints

$$ a=-A^2,{~}\gamma_1=-{ \alpha^2 \over 2},{~} \gamma_2=- \alpha
\beta, {~} \gamma_3=- \alpha A(N-1) \eqno(3) $$

where $ A, \alpha, \beta $ are arbitrary real constants,
then $N$ coordinates of the particles are $N$ real zeros of the
following function

$$ \phi (x,t) = \sum_{k=0}^N x^k[exp(t \hat T) \vec c(o)]_k \eqno(4)$$

where
$$ \hat T_{00}= -\beta,{~} \hat T_{02}= -a,{~} \hat T_{N-1N}=
-N \beta,{~} \hat T_{N-1N-2}= 2\alpha,{~} \hat T_{NN-1}= \alpha $$
$$ \hat T_{kk-1}=(N+1-k) \alpha,{~} \hat T_{kk+1}= -(k+1) \beta,{~}
\hat T_{kk+2}={ (k+1)(k+2) \over 2}a,$$
$$1 \le k \le N-2{~} \eqno(5)$$
and
$$c_{N-k}(0)=(-1)^k \sum_{1 \le {\lambda_1}<...< {\lambda_k}\le N}
x_{\lambda_1}(0)...x_{\lambda_k}(0),{~~~~~} c_N(0)=1.$$

Other elements of the matrix $ \hat T $ are zeros.
In these expressions $ x_i(0), i=1,...N$ are the initial
coordinates of $N$ particles.

\centerline{\bf 2. Two-parameter class of potentials.}

Further we apply the method briefly described to obtain
explicit solutions for the following two-parameter class of
potentials of the type (2)-(3). Consider the following constraints

$$ \beta=0,{~~~~~~~} \alpha=-A^2k \eqno(6) $$

where $k, A$ are arbitrary constants. The case $k=1$ was
considered in [6].Therefore we have two parameters $A$ and $k$ in the
potentials $ V(x),{~} W(x) $.  Constant $A$ represents the
intensity of pair interaction, so it can be called a coupling
constants for the system in question. Parameter $k$ defines
the ratio of the pair interaction intensity to the
intensity of the external field. At $k \rightarrow 0$
particles are influenced by the pair interaction only, and at
$k \rightarrow \infty$ we have no pair interaction. In this case
particles are influenced by the external field only.

\centerline{\bf 3. Two particles}

Here we apply the method to the simplest case of two particles.
For $N=2$ from (4)-(5) one gets $ \hat T^3=(-2a^3k^2)E$, where $E$ is
a unit matrix. In this case the row in (4) can be summed up and
the coordinates are the two solutions of the following equation

$$ \sum_{k=0}^2 { \left({x \over {-a(2k^2)^{1 \over 3}}} \right)}^k
S_k(\theta) {\bar T}^k \vec c(o)]_k =0 \eqno(7)$$

where $ a \bar T=\hat T,{~~} \theta= -at(2k^2)^{1/3},$

$$ S_0(\theta)={1 \over 3}(e^\theta+2e^{-\theta/2}cos({\sqrt{3}
\over 2}\theta)),$$
$$ S_1(\theta)={1 \over 3}
(e^\theta-2e^{-\theta/2}cos({\sqrt{3} \over 2}\theta+{ \pi \over
3})) $$
$$ S_2(\theta)={1 \over 3}(e^\theta-2e^{-\theta/2}cos({\sqrt{3}
\over 2}\theta-{ \pi \over 3}))$$

At $k= 1$ the asymptotic form of this equation for
arbitrary initial values is the following
$x^2-2^{1/3}x+2^{2/3}=0$.  Thus we come to the conclusion that
in some finite time interval $t_0$ the particles fall into the
singularity of the pair interaction. However at $k \ne 1$
for arbitrary initial values $x_1(0)$ and $x_2(0)$ there
exist such values of parameter $k$ that particles do not fall into
the singularity of the potential. For example, in case
$x_1(0)=-c, {~}x_2(0)=c$
particles do not fall into the singularity if the value of $k$
satisfies the following inequality

$${1  \over k^2} > c^2-1 $$

For arbitrary initial conditions the system does not collapse
under the condition that both the initial coordinates and $k$
satisfy the following inequality

$$
k^38x_1(0)x_2(0)(x_1(0)+x_2(0))+k^24(x_1(0)^2x_2(0)^2-$$
$$-(x_1(0)+x_2(0))-1)+
k4(x_1(0)+x_2(0))+4(x_1(0)x_2(0)-1)-(x_1(0)+x_2(0))^2<0$$

This inequality means that at some  $k$ the attraction of
the particles due to the pair potential is compensated by
the repelling due to the external field so that the
particles do not fall into the singularity in any finite
time interval.

\centerline{\bf 4. Three particles}

In this section we consider
$N=3$ system.  Under constraints (6) one can get
$ \hat T^4=0 $ , and the algebra of T-matrix becomes nilpotent.
Further we consider symmetrical initial conditions

$$ x_1(0)=-x_0,{~}x_2(0)=0, {~} x_3(0)=x_0 \eqno(8)$$

Suppose $ r=1/(-A^2t)$.  Then (8) has the following Cardano
form:

$$ y^3+uy+q=0, {~~~~~~~~} x=y-{ r^2(3+x_0^2r) \over 3Z }
\eqno(9) $$
where
$$ Z=r^3-x_0^2k^2r-k^2,{~~~} u={
{-r^4(x_0^4k^2+9r+3x_0^2r^2) } \over 3Z^2},{~~~}$$

$$ q={{k\sum_{m=0}^8 C_mr^m} \over Z^3}, $$
$$
C_0=k^4,{~~}C_1=3k^4x_0^2,{~~}C_2=3k^4x_0^4,{~~}C_3=k^2(k^2x_0^6-3),{~~}$$
$$C_4=-6k^2x_0^2,{~~}C_5=-3k^2x_0^4,{~~}C_6=4 -{2 \over 27 }k^2x_0^6,$$
$${~~}C_7= 4x_0^2,{~~} C_8={2 \over 3}x_0^4 $$

Solutions of (9) are determined by the value of the
discriminant $ Q=-108((u/3)^3+(q/2)^2) $ [8].  At $ 0 \le t < t_0 $
we thus have $ Q>0 $ and as a consequence three real roots being
the three coordinates of interacting particles:

$$ y_1=N+M,{~~~}y_{2,3}=-{{N+M} \over 2} \pm i {\sqrt 3}
{{N-M} \over 2}$$
where
$$ N=(-{q \over 2}+ \sqrt{-Q \over 108} )^{1 \over 3},{~~~}
   M=(-{q \over 2}- \sqrt{-Q \over 108})^{1 \over 3},$$
provided $MN=-u/3$.
At $t=t_0$ we have $Q=0$ , which leads to two equal coordinates of
two particles. In other words at this moment two of the particles
fall into the singularity of the potential of pair interaction.

For arbitrary initial conditions one can find from (3)--(5) the
asymptotic form of equation (9) to be (provided $ A\not=0 $ ) :

$$ (1-kx_1(0)x_2(0)x_3(0))(x^3-{1 \over k})=0{~}.$$

From this relation one can easily conclude that the system in
question collapses in finite time interval for any initial
conditions.

Consider a limiting procedure which enables us to find exact
expression for $t_0$.  Suppose $A=z,{~} k=1/z^2$. Suppose
$z \rightarrow 0$.  This limiting procedure corresponds to
infinitely weak pair interaction at finite external field.
 In this case it is easy to find $t_0$ explicitly.
Consider initial conditions of the form
$x_1(0)=x_1, x_2(0)=x_2, x_3(0)=0$. Without pair interaction the
third coordinate remains equal to zero. Two other solutions are the
following

$$ U(t,k,x_1,x_2)=t^2-{{12^{1 \over 3}(x_1+x_2)} \over
{kx_1x_2}}t+{12^{2 \over 3} \over {k^2x_1x_2}}$$

$$ x_{1,2}(t)={1 \over U(t,k,x_1,x_2)}{-12^{1 \over 3} \over
{k}}(t-{12^{1 \over 3} \over {kx_{2,1}}})$$

From these expressions one can get for $t_0$

$$ t_0 = \left({3 \over 2} \right)^{1 \over
3} \left|{{(x_1+x_2) \left(1-{{|x_1-x_2|} \over{|x_1+x_2|}} \right)}
\over {kx_1x_2}}\right|$$

Using the discriminant one can get for general initial conditions the
following upper estimate for  $ t_0$:

 $$ t_0  = {1 \over {4A^2(k+k^{-1/4})}} |((1-k) +
 9(x_1(0)x_2(0)+x_1(0)x_3(0)+x_2(0)x_3(0)) +$$
$$3(k-1)x_1(0)x_2(0)x_3(0)-21(k-1)(x_1(0)+x_2(0)+x_3(0)){1 \over
{kx_1(0)x_2(0)x_3(0)}}  | \eqno(10)$$

This estimate is valid for $ x_i(0) \not= 0, {~}i=1,2,3 $. It is
 easy to see that $t_0$ increases as  $A^2$ and/or $k$ decrease.
For some initial conditions there exists local maximum of
$t_0$ at finite $k>0$.  For example, for $ x_1(0)=-x_0,{~}x_2(0)=x_1,
 {~} x_3(0)=x_0$, if
$$x_1>{1  \over 129}{{24x_0^2(12+\sqrt{15})-43} \over{x_0^2+7}}$$

or
$$x_1<{1  \over 129}{{24x_0^2(12-\sqrt{15})-43} \over{x_0^2+7}}$$
there is a local maximum of $t_0$ at $k_0$.  We do not write down
the exact expression due to its complexity. A simple upper
estimate reads as follows

$$ k_0= {1 \over
{8W}} \left( 9(W-8x_0^2)+\sqrt{129(W^2-13x_0^2W+41x_0^4)} \right) $$
where $W=1+3x_1(7+x_0^2)$.

\centerline{\bf 5. Four particles}

For $N=4$ systems equations (4)-(5) lead to ${\hat
T}^5=18k^2a^3 {\hat T}^2$. In this case it is possible to sum up the
row in (4). After summation equation (7) turns into the following
expression

$$ \sum_{l=0}^2 {x^l} S_l(\theta) {\bar T}^l \vec c(o)]_l =0 $$

where $ a \bar T=\hat T,{~~} \theta= -at(2)^{1/3},$

$$ S_0(\theta)=1,{~~}
S_1(\theta)=(18)^{-1/3}k^{-2/3} \theta,$$
$$ S_2(\theta)={9 \over
12^{1/3}k^{4/3}}(e^\theta-2e^{-\theta/2}cos({\sqrt{3} \over
2}\theta-{ \pi \over 3})){~},{~~} S_3(\theta)={1 \over
 54k^{2}}(e^\theta+2e^{-\theta/2}cos({\sqrt{3} \over 2}\theta)- 3),$$
$$ S_4(\theta)={(18)^{-1/3} \over
54k^{8/3}}(e^\theta-2e^{-\theta/2}cos({\sqrt{3} \over 2}\theta+{ \pi
\over 3}) - 3 \theta) \eqno(13)$$

Therefore the four coordinates of interacting particles at any
given time are defined by the real roots of

 $$ \sum_{l=0}^4 R_l x^l = 0.  \eqno(14)$$

For initial conditions of the form
 $x_1(0)=-2x_0, x_2(0)=0,x_3(0)=x_0,x_4(0)=2x_0,$
coefficients in (14) have the form
$$R_0=24k^3x_0^3 S_3 -
8k^2x_0^2S_2 -x_0(S_1+30k^3S_4) +(1+12k^2S_3) {~~},$$
$$R_1=24k^3x_0^3 S_2 -
8kx_0^2(S_1+18k^2S_4) -x_0(1+30k^2S_3) +12kS_2$$
$$R_2=4x_0^3(3kS_1+54k^3S_4)
- 4x_0^2(1+18k^2S_3) -15kx_0S_2 +(6S_1+108k^2S_4) $$
$$R_3=4x_0^3(1+6k^2S_3) - 8kx_0^2S_2 -3x_0(S_1+10k^2S_4)
+12kS_3{~~}$$ $$R_4=-12kx_0^3S_2+4x_0^2(S_1+18k^2S_4) +15kx_0S_3
-6S_2 $$

At $t \rightarrow \infty$ the asymptotic form of (14) reads
as follows

$$ x^4 + k^{-1/3}(18)^{1/3}x^3 + {3 \over 2}k^{-2/3}(12)^{1/3}x^2 +
k^{-1}x - {1 \over 2}k^{-4/3}(18)^{1/3} = 0 $$

This equation has only two real roots. Thus we come to the
conclusion that for arbitrary $x_0$ two particles
fall into the singularity of the potential in finite time interval
$ t_0 $.

For arbitrary $x_i(0)$ and $k=1$ by reducing (14) to the form $x^4
+ ux^2 + qx + r =0$ and using the general expression for the
discriminant
[8] $ D= 16u^4r-4u^3q^2-128u^2q^2+144uq^2r-27q^2+256r^3$ ,
one can derive the following upper estimate for $ t_0$:
$$ t_0 = {{(18)^{1/3}} \over {4A^2}} {
{\sum_{j<l}^4 x_j^2(0)x_l^2(0)} \over {\sum_{j=1}^4 x_j^2(0)}}
\eqno(15) $$

One can conclude from (15) that the smaller the coupling
constant $A^2$ is, the greater $t_0$ is.

By now at $N>4$ such estimates have not been derived yet.
For example, at $N=5$ one can obtain from (4)-(5)
$$\hat R^6= \hat R^3+ bE, $$
where $\hat R= (2^{-1}(9)^{-1/3}k^{-2/3}) \hat T$
and $b=(5/2)^{2}3^{-4}$,{~} $E$--
identity matrix.  In this case we have
$$
S_0(\theta)=1+{{b \theta^6}\over 6!}+ \sum_{l=1}^\infty{{a_l
\theta^{9+3l}} \over {(9+3l)!}}, $$
where

$$ a_l=b\sum_{m=0}^{l-3}{{{(l+m)!} \over {(l-1)!m!}}b^m}. $$

Therefore summation of rows in (4) becomes more complicated.

\centerline {\bf 5.6 Summary}

We have considered Calogero-Moser systems in external fields under
special constraint (6). In fact we have considered
two-parameter subclass of potentials defined by  (1), (2).
For two-, three- and four-particle systems exact solutions have been
derived and analyzed. We have derived estimates for the collapse time
of the systems in question. It is shown that $N=2$ case is unique in
the sense that there are such values of the parameters in potentials
(8),(9), that the system has no finite collapse time $t_0$.
For $N \ge 3$ there are no such values.

 Explicit solutions obtained can be used for verification of
different approximate methods and for investigation of the systems
close to integrable ones. It is also possible that corresponding
quantum systems will possess especially simple behavior.

\vskip 0.2in
{\bf REFERENCES }

[1] Calogero F.// Lett.Nuovo Cim. 1975. {\bf 13}.P.411; J.Moser// Adv.
Math. 1975. {\bf 16}. P.197.

[2] Dittrich J., Inozemtsev V.I.// J. Phys. 1993. {\bf A20}. P.753.

[3] Olshanetsky M.A., Perelomov A.M.// Phys. Rep. 1983. {\bf 94}. P.312.

[4] Koprak Th.T., Wagner H.J.//J.Stat.Phys. 2000. {\bf 100}. P.779.

[5] Inozemtsev V.I., Meshcheryakov D.V.// Phys. Lett. 1984. {\bf A106}. P.105.

[6] Meshcheryakov D.V, Tverskoy V.B.// Moscow University Phys. Bull. 2000.
{\bf1}. P.66

[7] Korn G.A., Korn Th.M.// Mathematical Handbook. McGrow-Hill, New
York 1965.

[8] Feferman S.// The number systems. Foundation of algebra and
analysis.  Addison--Wesley, Reading 1964.
\bye